\documentclass[%
 rsi,
 amsmath,amssymb,
 reprint,%
longbibliography
]{revtex4-1}

\usepackage{graphicx}
\usepackage{dcolumn}
\usepackage{bm}

\usepackage[utf8]{inputenc}
\usepackage[T1]{fontenc}
\usepackage{mathptmx}
\usepackage{etoolbox}
\usepackage{xcolor}
\usepackage[hyperindex=true]{hyperref}

\definecolor{linkcol}{rgb}{0.2,0.2,0.6}

\hypersetup{bookmarksopen=true,
pdfmenubar=true, 
pdfhighlight=/O, 
colorlinks=true, 
pdfpagemode=None, 
pdffitwindow=true, 
linkcolor=linkcol, 
citecolor=linkcol, 
urlcolor=linkcol 
}

\makeatletter
\def\@email#1#2{%
 \endgroup
 \patchcmd{\titleblock@produce}
  {\frontmatter@RRAPformat}
  {\frontmatter@RRAPformat{\produce@RRAP{*#1\href{mailto:#2}{#2}}}\frontmatter@RRAPformat}
  {}{}
}%
\makeatother

\begin{document}

\preprint{AIP/123-QED}

\title{Bullet pressure-cell design for neutron scattering experiments with horizontal magnetic fields and dilution temperatures}
\author{Ellen Fogh$^*$}
\email{ellen.fogh@epfl.ch}
\affiliation{Laboratory for Quantum Magnetism, Institute of Physics, \'{E}cole Polytechnique F\'{e}d\'{e}rale de Lausanne (EPFL), CH-1015 Lausanne, Switzerland}
\author{Gaétan Giriat}
\affiliation{Laboratory for Quantum Magnetism, Institute of Physics, \'{E}cole Polytechnique F\'{e}d\'{e}rale de Lausanne (EPFL), CH-1015 Lausanne, Switzerland}
\author{Richard Gaal}%
\affiliation{Laboratory for Quantum Magnetism, Institute of Physics, \'{E}cole Polytechnique F\'{e}d\'{e}rale de Lausanne (EPFL), CH-1015 Lausanne, Switzerland}
\author{Oleksandr Prokhnenko}
\affiliation{Helmholtz-Zentrum Berlin f\"{u}r Materialien und Energie, D-14109 Berlin, Germany}
\author{Maciej Bartkowiak}
\affiliation{Helmholtz-Zentrum Berlin f\"{u}r Materialien und Energie, D-14109 Berlin, Germany}
\author{Luc Testa}
\affiliation{Laboratory for Quantum Magnetism, Institute of Physics, \'{E}cole Polytechnique F\'{e}d\'{e}rale de Lausanne (EPFL), CH-1015 Lausanne, Switzerland}
\author{Jana Pásztorová}
\affiliation{Laboratory for Quantum Magnetism, Institute of Physics, \'{E}cole Polytechnique F\'{e}d\'{e}rale de Lausanne (EPFL), CH-1015 Lausanne, Switzerland}
\author{Ekaterina Pomjakushina}
\affiliation{Laboratory for Multiscale Materials Experiments, PSI Center for Neutron and Muon Sciences, Forschungsstrasse 111, 5232 Villigen PSI, Switzerland}
\author{Yoshiya Uwatoko}
\affiliation{Institute for Solid State Physics (ISSP), University of Tokyo, Kashiwa, Chiba 277-8581, Japan}
\author{Hiroyuki Nojiri}
\affiliation{Institute for Materials Research, Tohoku University, Sendai 980-8577, Japan}
\author{Koji Munakata}
\affiliation{Neutron Science and Technology Center, Comprehensive Research Organization for Science and Society (CROSS), Tokai, Ibaraki 319-1106, Japan}
\author{Kazuhisa Kakurai}
\affiliation{Neutron Science and Technology Center, Comprehensive Research Organization for Science and Society (CROSS), Tokai, Ibaraki 319-1106, Japan}
\author{Henrik M. R{\o}nnow}
\affiliation{Laboratory for Quantum Magnetism, Institute of Physics, \'{E}cole Polytechnique F\'{e}d\'{e}rale de Lausanne (EPFL), CH-1015 Lausanne, Switzerland}

\date{\today}

\begin{abstract}

The simultaneous application of high magnetic fields and high pressures for controlling magnetic ground states is important for testing our understanding of many-body quantum theory. However, the implementation for neutron scattering experiments presents a technical challenge. To overcome this challenge we present an optimized pressure-cell design with a novel bullet shape, which is compatible with horizontal-field magnets, in particular the high-field magnet operating at the Helmholtz-Zentrum Berlin. The cell enabled neutron diffraction and spectroscopy measurements with the combination of three extreme conditions: high pressures, high magnetic fields, and dilution temperatures, simultaneously reaching $0.7\,\mathrm{GPa}$, $25.9\,\mathrm{T}$, and $200\,\mathrm{mK}$. Our results demonstrate the utility of informed material choices and the efficiency of finite-element analysis for future pressure-cell designs to be used in combination with magnetic fields and dilution temperatures for neutron scattering purposes.

\end{abstract}

\maketitle

\section{Introduction}

To map out phase diagrams and to characterize critical behavior close to quantum phase transitions it is essential to be able to control applied magnetic fields, hydrostatic pressures and temperatures simultaneously during an experiment. The Zeeman term in the spin Hamiltonian is exact and the magnetic field strength may be tuned accurately \textit{in situ}. Hydrostatic pressure is a means of manipulating interatomic bond lengths and angles and thereby alter magnetic interactions while preserving the chemical composition. Many quantum magnets require low temperatures to reach the interesting part of their phase diagrams \cite{sachdev2000,vasiliev2018}. Finally, neutron scattering is a unique technique for studying spin correlations directly \cite{neutronbook} and because neutrons penetrate matter easily compared to other scattering probes, such as X-rays or electrons, it is possible to have large equipment for controlling the sample environment. Therefore, neutron scattering experiments for studying quantum magnetic systems under applied magnetic field and pressures and at low temperatures constitute an extremely powerful tool.

\begin{table*}
	\caption{Overview of capabilities for combining high pressures, high magnetic fields, and low temperatures ($<10\,\mathrm{K}$) at different neutron scattering facilities (non-exhaustive list). The available sample space is also listed for each pressure cell. Note that the listed pressures are the values at room temperature. Figure~\ref{fig:facilities} represents these data in a graphical format.}
	\label{tab:facilities}
	\begin{ruledtabular}
		\begin{tabular}{l l c c c c}
			Facility 							& Cell type		& Pressure (GPa)	& Magnetic field (T)	& Temperature (K)	& Sample space (mm$^3$)\\
			\hline
			ISIS \cite{goodway, ISISwebpage}		& Clamp			& 1.75		& 9			& 1.8		& $390$\\
			ILL	\cite{ILLwebpage}				& P.-E.			& 6.0		& --			& 1.5		& $48$\\
												& Gas			& 0.7		& 6			& 1.5		& $1980$\\
												& McWhan (under repair)		& 3			& --			& 0.5		& $35$\\
			PSI \cite{khasanov2022, PSIwebpage} 	& Clamp			& 3.0		& 11			& 0.020		& $785$\\
												& P.-E.			& 5.5		& --			& 6			& $21$\\
			JRR-3 \cite{aso2007, kaneko2024, osakabe2010}	& Clamp	& 2.0		& --			& 0.8		& $7$\\
												& Hybrid	 anvil	& 10			& 10			& 3			& $0.2$\\
			JRR-3 / HFIR \cite{dissanayake2019}	& Cubic anvil	& 7.0		& --			& 3			& $1$\\
			J-PARC \cite{komatsu2015}			& Clamp			& 1.8		& --			& 0.1		& $25$\\
			IBR-2 \cite{kozlenko2018}			& Sapphire anvil	& 12			& --			& 4			& $0.1$\\
												& Diamond anvil	& 50			& --			& 4			& $0.01$\\
			LLB \cite{mirebeau2004,goncharenko2004}
				(until end of 2019)				& Sapphire anvil	& 12			& 7.5		& 0.1		& $0.1$\\
												& Diamond anvil	& 50			& --			& 1.4		& $0.01$\\
			HZB (until end of 2019)				& Bullet			& 1.0		& 25.9		& 0.200		& $450$
		\end{tabular}
	\end{ruledtabular}
\end{table*} 

\begin{figure}[t!]
	\includegraphics[width = \columnwidth]{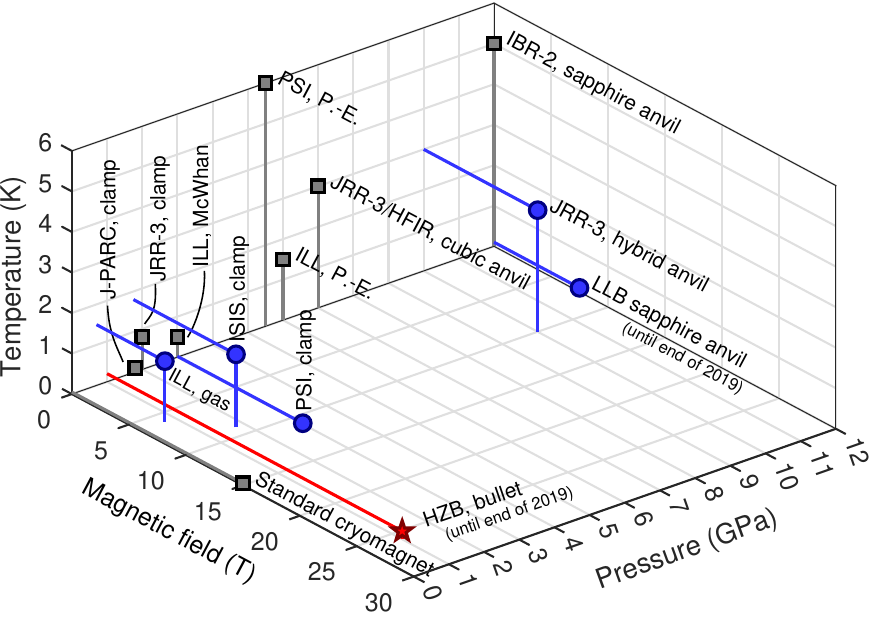}
	\caption{Representation of capabilities listed in Table~\ref{tab:facilities}. Black squares denote possibilities for combining low temperature with applied pressures or magnetic fields. Blue circles denote pressure cells with the possibility of combination with a magnetic field, i.e. the application of all three extremes. The red star is the bullet cell developed for the HFM/EXED instrument. Horizontal and vertical lines illustrate where the data points connect with the basal planes. Note that the 50 GPa ''Kurchatov-LLB'' diamond-anvil cell is omitted from this plot.}
	\label{fig:facilities}
\end{figure}

An overview of capabilities at different neutron scattering facilities is presented in Table~\ref{tab:facilities} and illustrated in Fig.~\ref{fig:facilities}. To our best knowledge the record for combining all three extremes of high pressure, high magnetic fields and low temperatures is $2.8\,\mathrm{GPa}$, $7\,\mathrm{T}$ and $0.1\,\mathrm{K}$ \cite{mirebeau2004}. Disregarding pressure for a moment, a combination of a dilution refrigerator and a standard vertical superconducting cryomagnet provides routinely $15\,\mathrm{T}$ and $50\,\mathrm{mK}$, which gives access to the $(H,T)$ phase diagram of many systems. Until the end of 2019 a High Field Magnet (HFM) facility for neutron scatttering was operated at the Extreme Environment Diffractometer (EXED) at the Helmholtz-Zentrum Berlin (HZB) \cite{prokhnenko2015,smeibidl2016,prokhnenko2016,prokhnenko2017}. The hybrid magnet consisted of a resistive and a superconducting coil, which together with a bespoke dilution refigerator, built in collaboration with the University of Birmingham, provided field strengths up to $25.9\,\mathrm{T}$ and temperatures down to $200\,\mathrm{mK}$ at the sample position. However, combining low temperatures with applied pressures is more challenging because of the power required to cool the pressure cell, which is typically a large piece of metal. Apart from aforementioned record, this has been achieved and documented in literature down to $1.4\,\mathrm{K}$ for pressures up to $43\,\mathrm{GPa}$ \cite{goncharenko2001}, down to $1.5\,\mathrm{K}$ for pressures up to $1.0\,\mathrm{GPa}$ \cite{kozlenko2008,ruegg2008}, down to $4.2\,\mathrm{K}$ for pressures up to $6.0\,\mathrm{GPa}$ \cite{terada2020,fogh2024b}, and down to $1.8\,\mathrm{K}$ for pressures up to $20.0\,\mathrm{GPa}$ \cite{klotz2016}, depending on the type of pressure cell and the type of measurement. Another challenge with regards to neutron scattering experiments in combination with high pressure is the rapidly decreasing sample space when demanding higher pressures. Therefore, these experiments are often feasible only for diffraction and when considering magnetic studies only for samples containing high-magnetic-moment ions typically found in rare earth compounds. Consequently, inelastic neutron studies of $S = 1/2$ systems under pressure are very few. 

\begin{figure*}
	\includegraphics[width  = \textwidth]{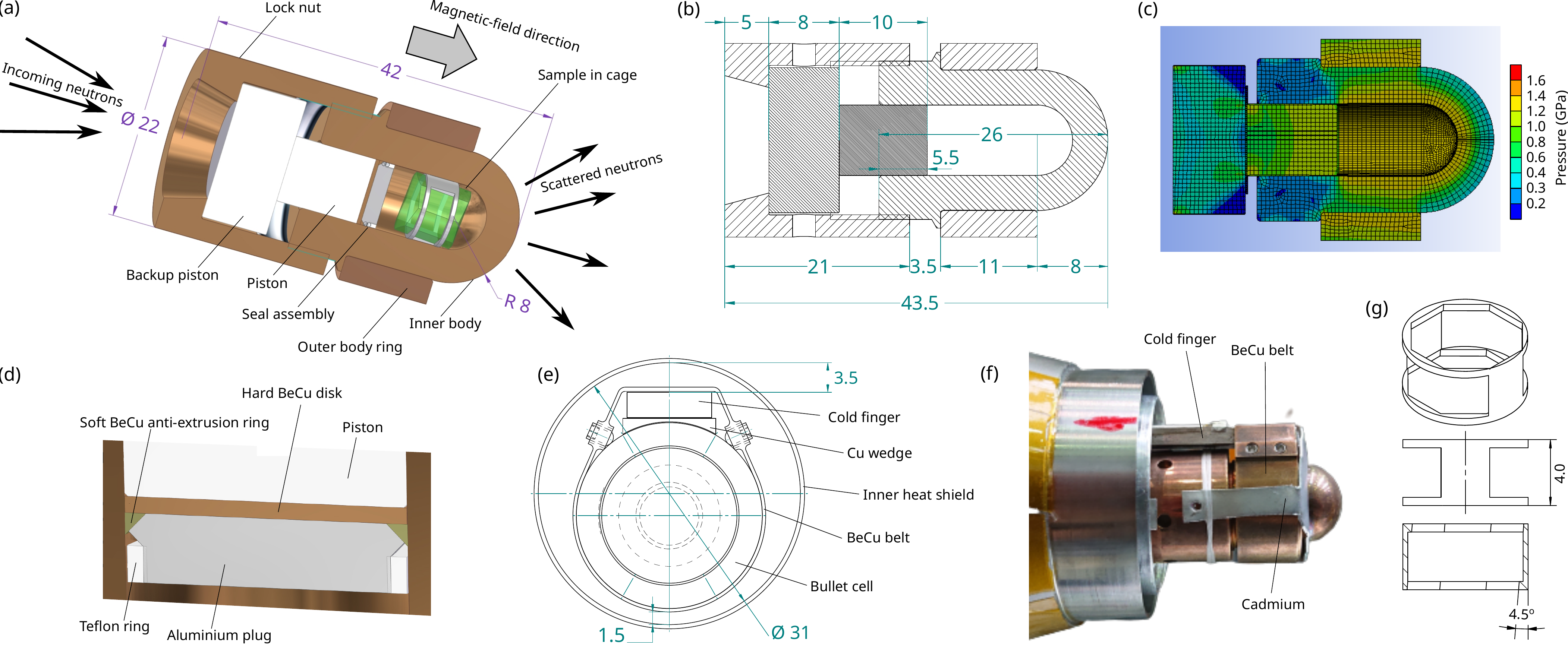}
	\caption{\textbf{Pressure cell design.} Dimensions in all panels are given in millimeters. (a) 3D representation and (b) schematic drawing of the bullet cell (side view). (c) Color contours showing the pressure distribution at $1.2\,\mathrm{GPa}$ in the sample space as simulated using finite-element analysis. (d) Illustration of the seal assembly. (e) Schematic drawing of the cell mounted on the cold finger of the dilution refrigerator (front view). (f) Photograph of the cell mounted on the cold finger before closing the heat shields (side view). (g) Schematic drawings of the sample cage.}
	\label{fig:bullet}
\end{figure*}

In this paper, we present a piston-cylinder pressure cell compatible with the dilution insert of the unique HFM/EXED instrument at the HZB. With an ultra-compact-cell design and optimized material choices, we succeeded in performing an inelastic neutron scattering experiment at $25.9\,\mathrm{T}$, $0.7\,\mathrm{GPa}$, and $200\,\mathrm{mK}$, and in this way probing spin excitations under these extreme conditions. The novel bullet design, meaning with a dome shape on the outgoing neutron beam side, allows the scattered signal to escape in a wide range of angles and ensures uniform attenuation. The concept opens an alternative path to explore for future pressure-cell engineering for neutron scattering purposes, in particular in combination with horizontal magnetic fields.

\section{Overview of apparatus}

The horizontal geometry and limited sample space of the $25.9\,\mathrm{T}$ magnet called for a highly specialized pressure-cell design. We used finite-element analysis as implemented in the software package ANSYS to simulate the pressure inside the cell and the strain on the components in order to optimize dimensions and material choices. The aim was to minimize the neutron background and attenuation, as well as to maximize the sample volume while obtaining $1.0\,\mathrm{GPa}$ at the sample position. Existing piston-cylinder cells typically consist of a vertical double-walled cylinder with pistons pushing from both top and bottom. In such cells, the neutrons travel through the cylinder walls in the transverse direction and not through the pistons. In the case of the spacially limited horizontal HFM, the neutrons were instead required to travel along the length of the cylinder. To minimize the neutron beam path through the pistons as well as to obtain uniform attenuation, we proposed a design consisting of a single piston assembly in the incoming beam path and a dome-shaped single wall at the other end. Due to this shape, it was named the ''bullet'' cell. A 3-dimensional (3D) representation and technical drawing of the bullet cell are shown in Fig.~\ref{fig:bullet}(a)-(b) and a color contour representation of the pressure distribution is shown in Fig.~\ref{fig:bullet}(c).

The outer diameter of the cell is $22\,\mathrm{mm}$ with a sample space diameter of $8\,\mathrm{mm}$. The sample space height is around $9\,\mathrm{mm}$ and the total length of the cell around $45\,\mathrm{mm}$ when closed. The exact dimensions when loaded depend on the piston displacement. The body of the cell is a double wall of hardened BeCu-25 with the outer support ring shrink fitted in place. The hardening of both body parts was carried out at 315$^{\circ}$C in a nitrogen atmosphere to prevent oxidation. The dome-shaped part of the cell consists of a single wall, which allows more neutrons to escape when scattered from the sample compared to a double wall. This shape also provides an optimal force distribution, such that a single wall can support the required pressure. On the incoming-beam side, the locknut has a conical opening to allow rotation of the cell with respect to the incoming neutron beam. Ceramic ZrO$_2$ (Coorstek Technox 3000) is the material choice for the piston in order to minimize the absorption of incoming neutrons when they travel through the piston to reach the sample. The ceramic is more transparent to neutrons than, for example, tungsten carbide, but it is brittle compared to metallic BeCu. Therefore the piston is the limiting part in this design with regards to the maximum achievable pressure.

The seal assembly [Fig.~\ref{fig:bullet}(d)] consists of four parts: a teflon ring, an aluminum plug, a soft BeCu anti-extrusion ring and a hardened BeCu disk. The assembly seals directly against the wall of the bore so no sample can is required. It is necessary to polish the bore inside prior to loading in order to remove any scratches that may cause leaks. This was performed using a lathe and SiC grinding paper down to Grit 4000 (US \#1200) and then diamond paste down to a grain size of $2\,\mathrm{\mu m}$. This procedure was repeated for each loading, which means that the bore gets slightly wider for each load and will eventually be too wide to be used with the same seal assembly and pistons. The anti-extrusion ring was covered with indium to create the initial seal. The slightly dome shaped BeCu disk inserted at the end serves to prevent the ceramic pistons from cracking by mediating the force more evenly over the surface.

All parts of the pressure cell, seal assembly and pistons were cleaned carefully using acetone and an ultrasound bath prior to loading. Afterwards, the sample was placed in the bore. The seal assembly was then prepared by squeezing the anti-extrusion ring with $120\,\mathrm{kg}$ on a hydraulic press using a steel ball with diameter $15\,\mathrm{mm}$ to achieve a slightly tighter fit inside the bore of the cell. The teflon ring was mounted on the aluminum plug. A deuterated 4:1 methanol-ethanol mixture was used as pressure medium and injected with a syringe to prevent air bubbles from becoming trapped under the sample. Deuterated methanol-ethanol was used to avoid the large incoherent neutron scattering associated with hydrogen \cite{sears1992}, which generates an undesired addition to the background. In order to allow sufficient piston displacement for the required pressure while engaging enough thread to sustain the load on the locknut, a pressure-medium level of approximately $4\,\mathrm{mm}$ as measured from the top of the bore was optimal. The seal was then placed by first inserting the aluminum plug with the teflon ring and using a brass plate to push it in evenly. The anti-extrusion ring was placed on top and also pushed in with the brass plate. The BeCu disk was placed in the bore with the top of the dome on the piston side. The assembly was engaged using a tungsten (W720) dummy piston and loaded to $200\,\mathrm{kg}$ for initial sealing. Finally, pressure was applied using the ceramic pistons while making sure that all parts were well centered, ensuring an even load and that the locknut would fit the threads once the desired piston displacement had been reached. Note that to minimize material, the locknut did not reach the thread of the cell body prior to loading. A holder fitted to the dome shape of the bullet cell body allowed us to carry out the loading on a standard hydraulic press. The load was increased slowly and continuously. Once the maximum load of $7200\,\mathrm{kg}$ was reached the cell was monitored for approximately $1\,\mathrm{h}$ to ensure that all components stayed in place and no leaks appeared. The locknut was tightened 3 times to $7000\,\mathrm{kg}$, with the load topped up to $7200\,\mathrm{kg}$ between each tightening. The duration of the press release was similar to the loading process while keeping track of the piston displacement in both directions. The loaded cell had a total mass of approximately $100\,\mathrm{g}$.

The cell was mounted on the cold finger of the dilution refrigerator for the HFM by using a BeCu belt of thickness $0.3\,\mathrm{mm}$, as shown in Figs. \ref{fig:bullet}(e)-(f). To adapt from the flat surface of the cold finger to the rounded shape of the pressure cell, as well as to ensure thermal contact, a Cu wedge was inserted between the two surfaces. The wedge was gold plated on the curved side. The belt was tightened by engaging screws on either side. Cadmium masks were added on both the incoming and scattered beam sides to reduce the amount of illuminated cell material and to prevent neutrons scattered off the cell from reaching the detector.

\begin{figure}
	\includegraphics[width = \columnwidth]{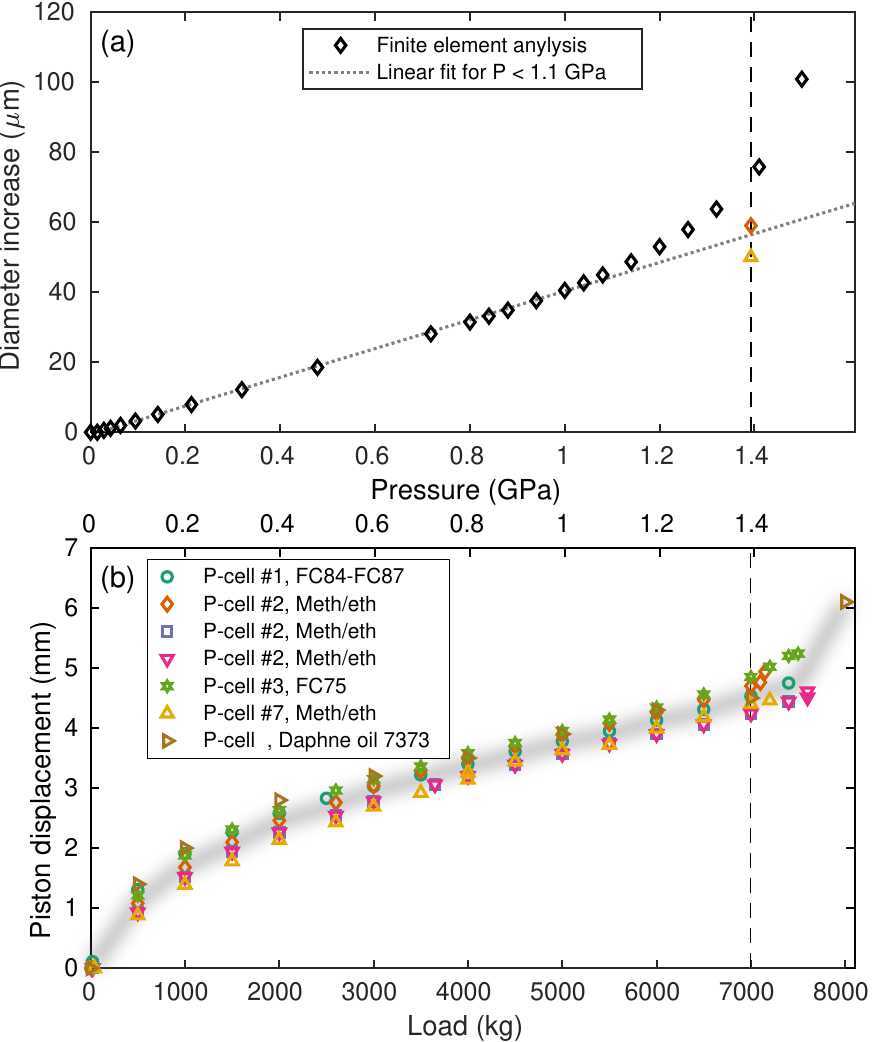}
	\caption{\textbf{Pressure cell loading.} (a) Increase of the cell diameter as a function of applied pressure as simulated in ANSYS with a linear fit for $P < 1.1\,\mathrm{GPa}$. The red diamond and yellow triangle are experimentally measured increases of the diameter at an applied load of $7000\,\mathrm{kg}$. (b) Piston displacement shown as a function of the applied load for 7 individual loadings using different copies of the cell as well as different pressure media. The dashed line indicates a load of $7000\,\mathrm{kg}$ where in the experiment the loading curve deviates from linear behavior. The gray shading is a guide to the eye.}
	\label{fig:loading}
\end{figure}

Figure~\ref{fig:loading}(a) shows the simulated diameter increase of the pressure cell as a function of the applied pressure. The curve deviates from linear behavior above around $1.1\,\mathrm{GPa}$. The corresponding measured piston displacement as a function of applied load is shown for 7 individual loadings in Fig.~\ref{fig:loading}(b). The loading curves are remarkably similar, regardless of the cell number and pressure medium. The linear behavior ends around $7000\,\mathrm{kg}$, which corresponds to $1.4\,\mathrm{GPa}$ at the sample position assuming no friction. Noting that the linear behavior of the piston displacement starts only around $1500\,\mathrm{kg}$, and allowing for some of the applied force being lost to friction, the simulation and experiment can be said to agree with each other. The fact that the loading curve is initially non-linear is explained by air trapped in the system, which has a much higher compressibility compared to the any of the pressure media used in our tests. Beyond a load of $7000\,\mathrm{kg}$, the BeCu deforms plastically. Especially the dome part of the cell with its single wall expands irregularly and does not revert to original size and shape upon depressurization. Note that upon cooling there is a substantial pressure loss of 30-50\% (pressure determination at $200\,\mathrm{mK}$ is treated in Section IV). Furthermore, once loaded the pressure was found to be leaking slowly (half life around 10 days with pressure fully lost after a month). This was observed by monitoring the decrease of the cell diameter over time. Consequently one needs to load the cell immediately before the experiment and cool it down without delay. Once the pressure medium is frozen, the pressure is conserved.

\section{Experiment}

The material of choice for testing the bullet pressure-cell design is the quantum magnet SrCu$_2$(BO$_3$)$_2$, which represents a physical realization \cite{miyahara1999} of what is known as the Shastry-Sutherland model (SSM) \cite{shastry1981}. The magnetic lattice consists of pairs of Cu$^{2+}$ ($S = 1/2$) ions (dimers) arranged orthogonally to all neighboring pairs. The superexchange interactions are $J$ within the dimer, and $J'$ between dimers. At ambient conditions, SrCu$_2$(BO$_3$)$_2$ is well described by the SSM with $J'/J \approx 0.63$ \cite{miyahara1999,kageyama1999,kageyama2000b,kageyama2000,gaulin2004,kakurai2005}, and the ground state of the system is a product of singlets on the dimers \cite{koga2000,chung2001,lauchli2002,corboz2013,nakano2018,xi2023}. Upon applying a magnetic field, a series of magnetization plateaux is observed \cite{onizuka2000,takigawa2013,matsuda2013,shi2022,nomura2023} with the first of these occuring at $27\,\mathrm{T}$. Here, the magnetization is 1/8 of its saturation value and a spin superstructure is predicted \cite{corboz2014}. Torque measurements and numerical calculations indicate that applying pressure to SrCu$_2$(BO$_3$)$_2$ suppresses the transition field, bringing it below $25.9\,\mathrm{T}$ \cite{haravifard2016,schneider2016,shi2022} and thus enabling a neutron scattering study of the $1/8$ magnetization plateau in SrCu$_2$(BO$_3$)$_2$ by using a combination of the bullet cell and the HFM/EXED facility.

The HFM allowed for direct probing of the magnetic structure and dynamics up to $25.9\,\mathrm{T}$ static magnetic fields, the bulllet cell provided a hydrostatic pressure of $0.7\,\mathrm{GPa}$, and the dilution insert took the system to a temperature of $200\,\mathrm{mK}$. The magnet had $30^{\circ}$ conical opening angles, which in combination with magnet rotation of $\omega \in \left[-11.8^{\circ},+2^{\circ}\right]$ from the incident neutron beam and the time-of-flight (ToF) technique implemented on EXED, gave access to a substantial region in momentum transfer, ${\bf Q} = (q_h,q_k,q_l)$, and energy transfer, $E$. A schematic representation of the setup is shown in Fig.~\ref{fig:diffraction}(a). For detailed descriptions and illustrations of this truly unique instrument and magnet we refer to Refs. \onlinecite{prokhnenko2015,prokhnenko2016,prokhnenko2017,smeibidl2016}. Three different chopper settings produced a wavelength band of $0.7-7.2\,$\AA~for diffraction and incoming energies $E_i = 4$ and $8\,\mathrm{meV}$ for inelastic measurements. A general characterization study of the bullet cell was performed by varying the magnet rotation angle. For collecting magnetic diffraction data we chose magnet rotation angle $\omega = 0$ and for inelastic measurements we positioned the magnet at $\omega = -6^{\circ}$. The latter angle was chosen as a compromise between access to a larger portion of reciprocal space and optimal neutron transmission through the pressure cell. Only forward scattering detectors were used in this experiment.

A single crystal of SrCu$_2$(BO$_3$)$_2$ grown by the floating-zone method \cite{kageyama1999c,jorge2004}, was cut with octagonal cross section perpendicular to the ${\bf c}$-direction. This shape was chosen as a compromise between maximizing the sample volume in the cylindrical sample space and still being practical to cut. The resulting sample had a mass of $0.8\,\mathrm{g}$ and was mounted in an aluminum cage as shown in Fig.~\ref{fig:bullet}(a) and Fig. \ref{fig:bullet}(g). The orientation of the SrCu$_2$(BO$_3$)$_2$ crystal was chosen to optimize the coverage in $({\bf Q}, E)$. Multiple scenarios for sample and magnet orientations were simulated using the software packages EXEQ and InEXEQ, developed specfically for experiment planning at the HFM/EXED facility \cite{bartkowiak2020}. Theoretical calculations predict a spin texture with ordering vector in the $(q_h,q_k,0)$-plane for the $1/8$ magnetization plateau \cite{corboz2014}. To observe this, the ${\bf c}$-direction of the crystal had to be tilted in the horizontal plane with respect to the magnetic field direction and with an optimal tilt angle of $4.5^{\circ}$. The sample cage was precut to obtain this tilt angle [Fig. \ref{fig:bullet}(g)]. The resulting sample orientation was determined by tracking nuclear Bragg peak positions on the detector upon magnet rotation.

\section{Results}

\begin{figure}
	\includegraphics[width = \columnwidth]{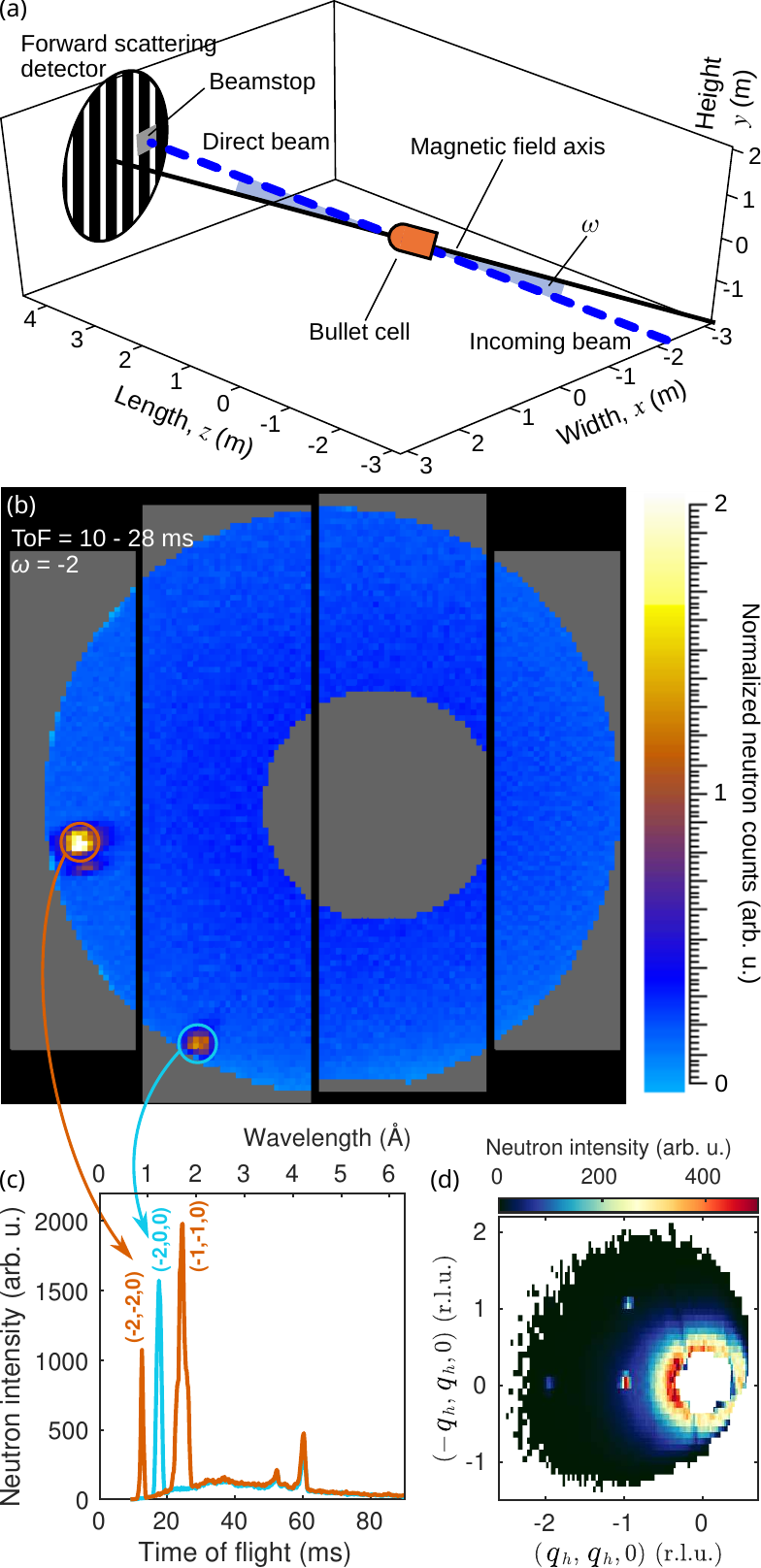}
	\caption{{\bf Experimental setup and nuclear Bragg scattering.} (a) Schematic representation of the instrument geometry, showing the magnetic-field axis (black line), the incoming-neutron-beam direction (dashed blue line), and the positions of the bullet cell and the detector. This image is based on simulations using the software package InEXEQ \cite{bartkowiak2020} (b) Detector image for magnet rotation $\omega = -2^{\circ}$ integrated over a ToF window of $10-28\,\mathrm{ms}$. Two high-intensity spots are clearly visible in the lower left corner of the detector. The upper spot corresponds to the nuclear Bragg peaks $(-2,-2,0)$ and $(-1,-1,0)$ and the lower to $(-2,0,0)$. 
	(c) 	Integrated neutron intensity at the detector positions of the two high-intensity spots shown as a function of neutron ToF. (d) Reciprocal space coverage corresponding to the detector image in panel (b).}
	\label{fig:diffraction}
\end{figure}

We start by examining the characterization results with a representative zero-field detector image shown in Fig.~\ref{fig:diffraction}(b). The corresponding ToF intensity profiles for the observed nuclear Bragg peaks and reciprocal coverage are shown in Fig.~\ref{fig:diffraction}(c)-(d). We point out that the irregular shape of the peaks in Fig. \ref{fig:diffraction}(c) are mainly due to intensity being spread over multiple detector tubes and to a lesser extent a result of sample mosaicity. Since the detector is constructed of arrays of vertical detector tubes with gaps in between some neutrons are not detected, but escape in between the tubes. This also explains why the shape of $(-2,-2,0)$ appears more regular compared to $(-1,-1,0)$ since it is spread over a smaller number of tubes.

All nuclear signals were observed for flight times below $40\,\mathrm{ms}$. At larger flight times, around $52$ and $60\,\mathrm{ms}$, two additional peaks were observed. These are of spurious origin, as we demonstrate further below. To estimate the overall signal-to-noise performance of the bullet cell we did the following for each magnet rotation angle: (i) the Bragg peak intensities were determined by numerical integration over the relevant detector pixels and ToF intervals like in Fig.~\ref{fig:diffraction}(b)-(c), (ii) the background signals were estimatated by integrating parts of the detector for similarly located regions but away from any Bragg-peak intensity (not shown), and (iii) the signal-to-noise ratio was evaluated. The maximum signal-to-noise ratio obtained was around 8 times lower with the bullet cell in the beam compared to the ambient-pressure experiment performed without a pressure cell \cite{fogh2024}.

\begin{figure}
	\includegraphics[width = \columnwidth]{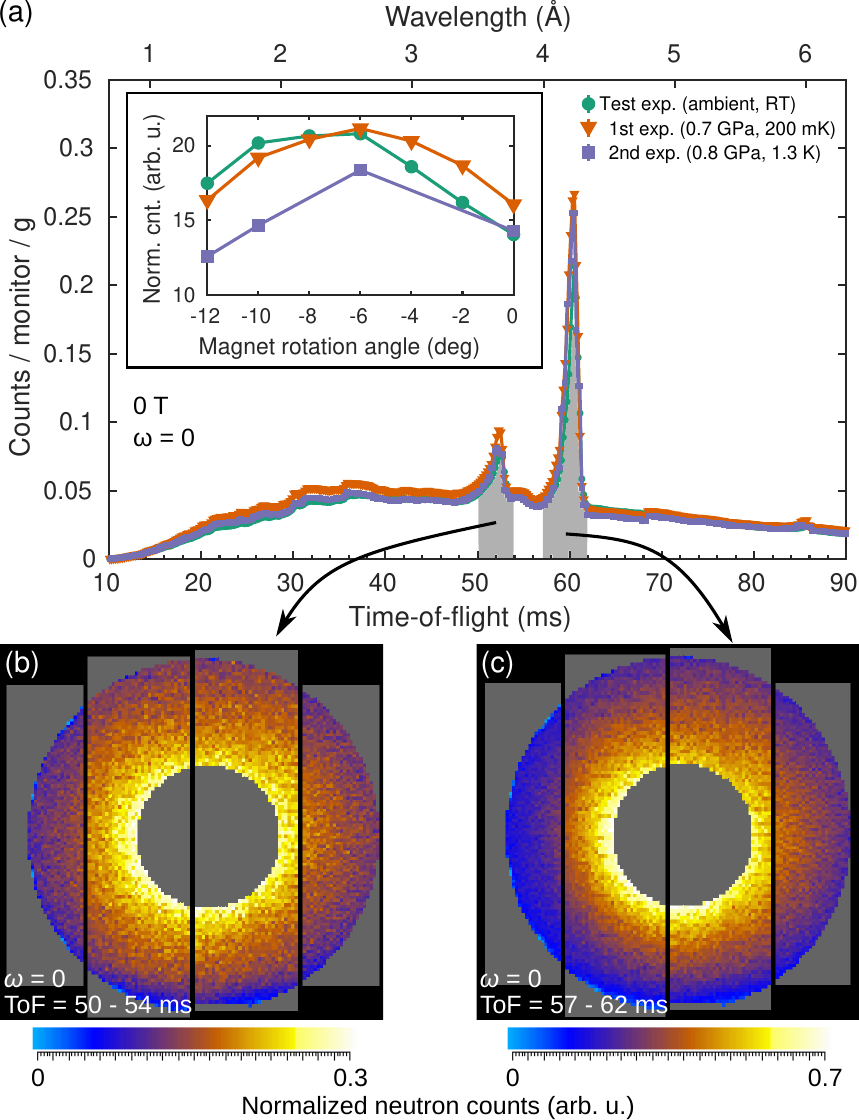}	
	\caption{{\bf Background from the bullet cell.} (a) Neutron intensity shown as a function of ToF for three different pressure cells mounted on EXED and with no magnet rotation ($\omega = 0$). The neutron counts were summed over the entire detector excluding regions with nuclear Bragg peaks and normalized with respect to the sample mass. In the test experiment, the pressure cell was not loaded (but contained a sample) and the experiment was performed at room temperature (RT). The inset shows total neutron counts as integrated over all flight times and as a function of magnet rotation angle. The transmission has a maximum around $\omega = -6^{\circ}$. 
	(b)-(c) Detector images for the two spurions around neutron flight times of $52$ and $60\,\mathrm{ms}$ corresponding to the shaded areas in panel (a). The elevated level of neutron counts is distributed evenly throughout the detector in both cases.}
	\label{fig:background}
\end{figure}

The total detector counts measured as a function of the neutron ToF are shown in Fig.~\ref{fig:background}(a) for three different pressure cells and samples. Regions of the detector containing nuclear Bragg peak intensity were excluded and hence the curves in Fig.~\ref{fig:background}(a) represent the background, which is due largely to the pressure cell but also due to the cryostat and magnet. There are two prominent features at $52$ and $60\,\mathrm{ms}$ on top of a very broad structure with maximum around $35\,\mathrm{ms}$. These features are also clearly present in Fig.~\ref{fig:diffraction}(b). The intensity around the peaks has no structure in $\bf Q$ [Fig.~\ref{fig:background}(b)-(c)] and are believed to originate from multiple scattering involving the BeCu pressure cell and one of the upstream aluminum windows. The spurious signal is likely enhanced by neutron transmission peaks around these wavelengths (3.6 and 4.1 Å) \cite{pressurebook}. For potential future experiments at other neutron facilities as well as for future cell designs we will consider the approach in Ref. \cite{ma2024} using the Union framework \cite{bertelsen2017} in the McStas Monte Carlo simulation package for neutron beamlines \cite{willendrup2020,willendrup2021} to identify possible spurions as well as ways to avoid them in experiments. Returning to the current cell design, the transmission is optimal around magnet rotation angle $\omega = -6^{\circ}$, which is why this particular angle was chosen for the inelastic neutron data collection. The small discrepancies between the curves in Fig.~\ref{fig:background}(a) (main panel and inset) are explained by slight differences in the cell orientation with respect to the incoming beam, which occur when mounting the cell on the cold finger. Note also that the data set at 1.3 K in Fig. \ref{fig:background}(a) was collected using a different cryostat which explains the overall lower background counts.

\begin{figure}
	\includegraphics[width = \columnwidth]{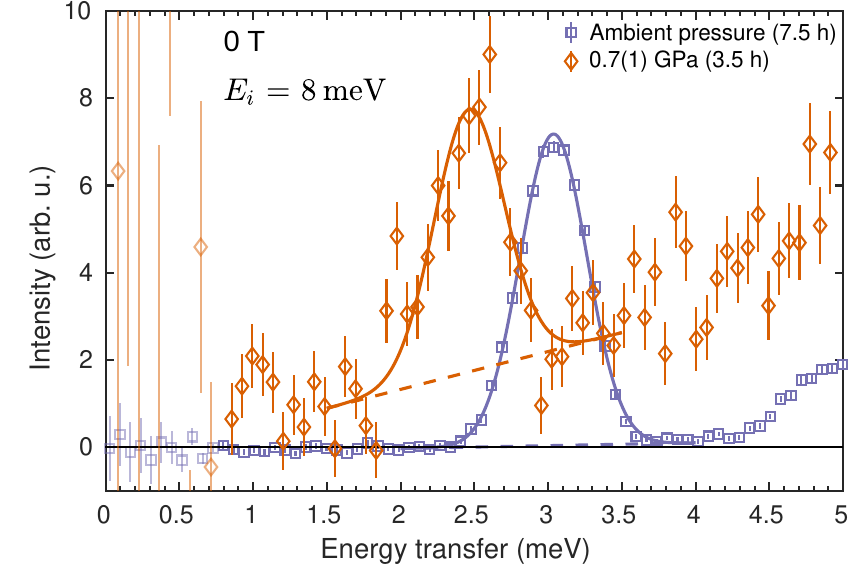}
	\caption{Background-subtracted neutron intensity shown as a function of energy transfer at ambient and at finite pressure. At ambient pressure, the first triplet excitation is found around $3.0\,\mathrm{meV}$ and is shifted to lower energies upon applying pressure. The data are scaled such that the integrated intensities of the Gaussian fits (solid lines) are equal. The data points shown in lighter colors represent the limit of our resolution close to the elastic line. The ambient-pressure data are reproduced from Ref. \onlinecite{fogh2024}.}
	\label{fig:tripletshift}
\end{figure}

The pressure was determined by the energy shift of the first triplet excitation in SrCu$_2$(BO$_3$)$_2$ at base temperature [Fig.~\ref{fig:tripletshift}]. The ambient-pressure reference was measured in a separate experiment without a pressure cell and a full treatment of these data is published elsewhere \cite{fogh2024}. The integration ranges in Fig.~\ref{fig:tripletshift} were $q_h = [-0.95,0.55]$, $q_k = [-1.05,0.45]$, $q_l = [-0.9,1.3]$ for the curve at finite pressure and $q_h = [-1.25,0.25]$, $q_k = [-0.75,0.75]$, $q_l = [-0.9,1.3]$ for the data at ambient pressure. The dispersionless nature of the triplet excitation in SrCu$_2$(BO$_3$)$_2$ renders a direct comparison between the mode positions in such two regions of $\bf Q$ a valid approach. The large integration ranges were required to improve the neutron counting statistics. The resulting spectra were fitted to Gaussian profiles with a sloping background and the fitted center positions yielded a shift of $\Delta E = -0.57(2)\,\mathrm{meV}$. Using the linear relation, $P = \frac{-\Delta E}{0.080(4)} \frac{\mathrm{GPa}}{\mathrm{meV}}$, between pressure and peak position given in Ref. \onlinecite{zayed2014} we obtain a pressure of $P = 0.7(1)\,\mathrm{GPa}$. At this pressure, the transition field to the $1/8$ magnetization plateau is predicted to be $25\,\mathrm{T}$ \cite{schneider2016,haravifard2016}, which means that it is sufficiently suppressed to be probed at the HFM/EXED facility.

\begin{figure*}
	\centering
	\includegraphics[width = \textwidth]{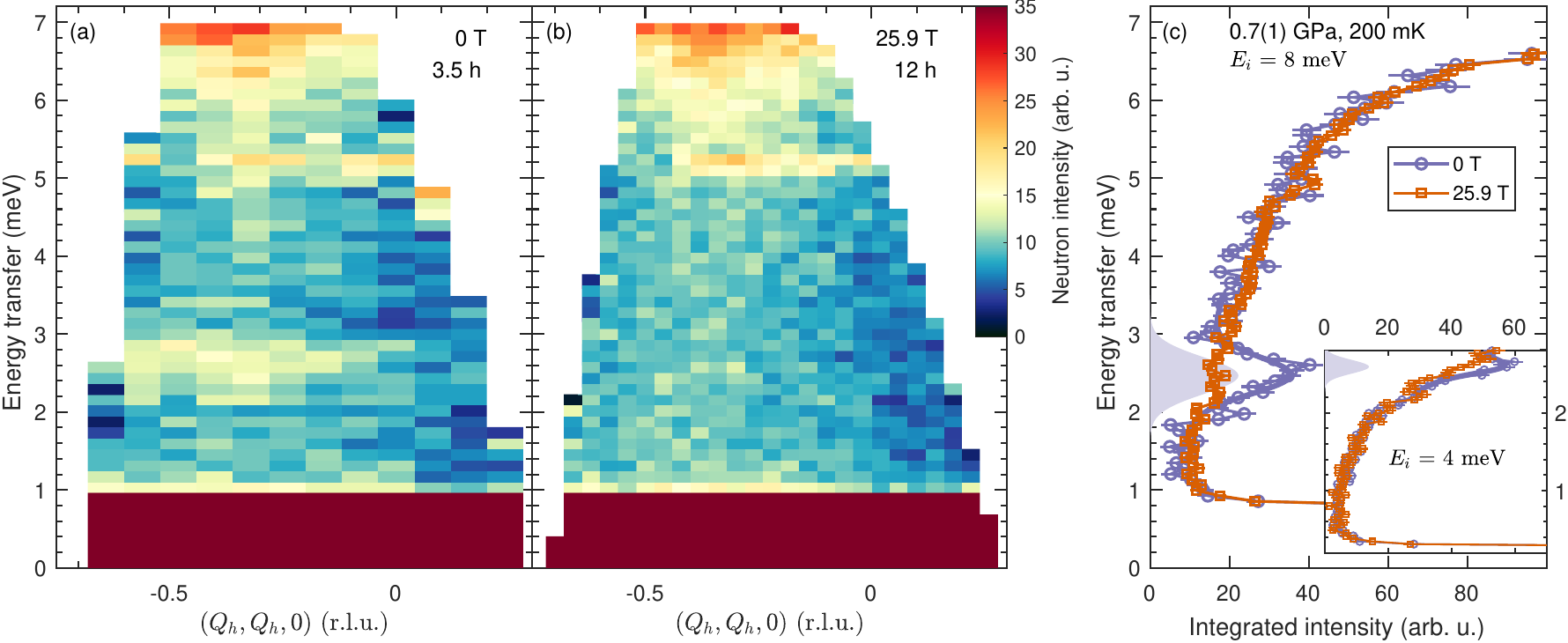}
	\caption{{\bf Inelastic neutron scattering spectra.} Neutron intensity at $0.7(1)\,\mathrm{GPa}$ and $200\,\mathrm{mK}$ shown as a function of energy transfer and for $(q_h,q_h,0)$ at (a) zero field and (b) $25.9\,\mathrm{T}$ measured with incoming energy $E_i = 8\,\mathrm{meV}$. The neutron intensity was integrated over $\pm0.5\,\mathrm{r.l.u.}$ perpendicular to $(q_h,q_h,0)$ and over the range $q_l = [-0.9,1.3]$. (c) Neutron intensity integrated over the entire available volume of reciprocal space shown as a function of energy transfer for data collected with $E_i = 8\,\mathrm{meV}$ for zero field (blue circles) and $25.9\,\mathrm{T}$ (red squares). The inset shows data with $E_i = 4\,\mathrm{meV}$. The triplet excitation is fitted to a Gaussian as described in the text and shown with blue shading.}
	\label{fig:INSresults}
\end{figure*}

We now turn to finite magnetic fields where extensive diffraction data were collected at $20\,\mathrm{T}$ (34~h), $25\,\mathrm{T}$ (35~h) and $25.9\,\mathrm{T}$ (22~h). No magnetic-field-induced intensity was observed and we discuss the possible reasons for this further on. The inelastic signal at zero field and at $25.9\,\mathrm{T}$ is compared in Fig.~\ref{fig:INSresults}(a)-(b). At zero field, the triplet excitation is located around $2.5\,\mathrm{meV}$ and it is almost dispersionless as expected. At $25.9\,\mathrm{T}$, the triplet excitation is no longer observed and the spectrum is almost featureless. As a general observation in both cases, the neutron intensity is larger for larger $|{\bf Q}|$. Figure~\ref{fig:INSresults}(c) shows the integrated neutron counts as a function of energy for the ranges $q_h = [-0.95,0.55]$, $q_k = [-1.05,0.45]$, $q_l = [-0.9,1.3]$ for measurements with incoming energy $E_i = 8\,\mathrm{meV}$ and $q_h = [-0.65,0.35]$, $q_k = [-0.7,0.3]$ and $q_l = [-0.65,0.75]$ for measurements with incoming energy $E_i = 4\,\mathrm{meV}$. The scattering intensity at $25.9\,\mathrm{T}$ is smeared out, which results in a slightly higher overall count rate but with no clear features compared to the zero-field spectrum. The resolution is improved when collecting data using $E_i = 4\,\mathrm{meV}$, as shown in the inset of Fig.~\ref{fig:INSresults}(c). Here, at zero field, the triplet mode is located at the edge of the accessible energy range and as a consequence its position is not well-determined. Nevertheless, it is clear that no new modes could be observed below the triplet at high fields.

We point out that the geometry of the HFM/EXED setup dictates that different values in $E$ correspond to different values in $q_l$ \cite{fogh2024,allenspach2022}. This is not a problem in the particular case of SrCu$_2$(BO$_3$)$_2$ where the excitations are highly localized in all directions (and therefore show very little dispersion). In a system with more significant dispersion the approach of integrating over such a large volume of reciprocal space to improve the neutron statistics would be invalid.

We also note the spurious signal around $5.1\,\mathrm{meV}$ in Fig.~\ref{fig:INSresults}(a)-(b). This signal does not come from the pressure cell, as it was also seen in the ambient-pressure experiment \cite{fogh2024} but is most likely also not from the sample itself because the spurion position is independent of magnetic field and pressure. In the data measured with incoming energy $E_i = 4\,\mathrm{meV}$, a corresponding spurion appears around $2.1\,\mathrm{meV}$. In the $\bf Q$-integrated data shown in Fig.~\ref{fig:INSresults}(c), a Gaussian with a linear background was fitted to the zero-field data to describe the spurious signal and subsequently subtracted.

\section{Discussion}

Our work underlines the challenge of performing neutron scattering experiments under the simultaneous application of several extreme conditions. We discuss below the performance of the bullet cell, interpret the results and suggest improvements for future designs. We point out that the discontinued operation of the BER-II research reactor at the HZB impeded thorough testing of the bullet cell using well-known samples as well as empty-cell measurements. This limited available beamtime meant that we aimed directly for doing a real experiment which in turn carries a higher risk due to more unknown variables. Nevertheless, we succeeded in measuring an inelastic neutron scattering signal under conditions of $25.9\,\mathrm{T}$, $0.7(1)\,\mathrm{GPa}$ and $200\,\mathrm{mK}$.

Because of the stiffness of the SrCu$_2$(BO$_3$)$_2$ structure ($<1$ \% reduction of lattice parameters in the $(a,b)$-plane at 5.5 GPa \cite{loa2005,haravifard2014} our pressure determination relies solely on the position of the triplet excitation of SrCu$_2$(BO$_3$)$_2$ in zero field together with its known evolution with pressure \cite{zayed2014}. This is the main weakness of the setup and gives rise to a high uncertainty in the pressure determination. With the $1/8$ plateau extending for only a narrow field interval (around $1\,\mathrm{T}$ wide) at a given pressure \cite{haravifard2016,levy2008,shi2022}, it is easy to overshoot or undershoot with the field. Below the $1/8$ plateau resides a spin-nematic phase \cite{fogh2024}, which is expected to be the ground state of the system at $25.9\,\mathrm{T}$ for pressures below $0.5\,\mathrm{GPa}$ \cite{schneider2016}. The spectrum in the spin-nematic phase was previously characterized both experimentally and theoretically \cite{fogh2024}, and does not ressemble the results in Figs.~\ref{fig:INSresults}(b)-(c). On the other hand, if the pressure is higher than estimated, then at $25.9\,\mathrm{T}$ the system could already be on the next plateau where the magnetization is $2/15$ of the saturated moment. Future pressure cell designs for similar setups should include an independent pressure gauge such as Pb or NaCl for more accurate pressure determination and extensive offline testing should be undertaken using e.g. ruby fluorescence \cite{mao1986}. Other general improvements to consider for the bullet-cell design is strengthening the piston by using e.g. Zr-toughened alumina which also has high transparency to neutrons. Higher load-bearing capabilities may also be reached by adding a metallic binder around the cylinder of the piston. Finally, to address the issue of pressure loss upon cooling, incorporating a spring mechanism into the design could help maintain the force while cooling. This approach was succesfully employed previously \cite{haberl2018}.

Whereas nuclear Bragg peaks from the sample were clearly visible, no magnetic-field-induced changes were observed in the neutron diffraction signal. This is expected in the spin-nematic phase at $20\,\mathrm{T}$, where there is no long-range order. However, magnetic order is predicted at higher fields, i.e. both on the $1/8$ and $2/15$ plateaux \cite{corboz2014} but we observed no magnetic Bragg peaks or diffuse magnetic intensity. There are a number of reasons for why such signal may not appear: (1) Truly two-dimentional order leads to rods of scattering. In the experiment, we bisect only a finite volume of the rods and do not capture the full scattering intensity. Consequently the magnetic signal could be too weak to be observed in this experiment. (2) We rely on model predictions for the choice of $\bf Q$ range. If the model is wrong, we could simply be looking in the wrong part of reciprocal space. (3) The energy scale of the excitations on the plateau is likely lower compared to at zero field, which means that the temperature scale also decreases. Therefore, $200\,\mathrm{mK}$ might not be cold enough to enter the plateau.

In general, when magnetic fields stronger than around $16\,\mathrm{T}$ are required, the only option currently available for neutron scattering experiments is that of pulsed magnetic fields. With this method, field strengths up to $40\,\mathrm{T}$ may be reached \cite{duc2018,fogh2020,gazizulina2021,holm-janas2024}, but only for a duration of a few milli seconds, which means a very short integrated measuring time. Consequently, this technique has to date only been suitable for diffraction studies where, as a rule, signals are orders of magnitudes stronger than inelastic signals. The pulsed-field path is therefore limited in applications and combinations with pressures and dilution temperatures have not been attempted. Conversely, the results presented here demonstrate the feasibility of pressure-cell engineering for horizontal magnets and may prove relevant for next-generation static magnets based on high-$T_C$ superconducting ceramic materials and delivering beyond $16\,\mathrm{T}$ \cite{markiewicz2012,bai2020}. Moreover, our experiment was carried out on a medium-flux reactor but the future European Spallation Source will provide orders of magnitude higher neutron flux which render neutron scattering experiments using a similar combination of extreme conditions for studying quantum magnetic systems promising. Until such time, the bullet-type pressure cells have potential for use with existing static horizontal-field magnets at various neutron facilities, in particular for combined pressure and field studies of spin textures with long periods such as skyrmions \cite{muhlbauer2009}, vortex lattices \cite{xie2024} and spin spirals \cite{romaguera2024}.

\section{Conclusions}

We designed and constructed a pressure cell for neutron scattering experiments in combination with applied horizontal magnetic fields. Its bullet shape and optimized material choices enabled collection of inelastic neutron scattering and neutron diffraction data at $0.7(1)\,\mathrm{GPa}$, $25.9\,\mathrm{T}$, and $200\,\mathrm{mK}$. Under these conditions it was possible to investigate the $1/8$ magnetization plateau in the frustrated quantum magnet SrCu$_2$(BO$_3$)$_2$. Our work demonstrates the capabilities and future directions for combining the three extremes of high pressures, high magnetic fields and low temperatures for state-of-the-art neutron scattering experiments in the field of quantum magnetism.

\begin{acknowledgments}
This work was funded by the European Research Council through the Synergy network HERO (Grant No. 810451) and by the Swiss National Science Foundation through Project Grant No.~188648.
We would like to extend our gratitude to the team running the High Magnet Facility at the Helmholtz-Zentrum Berlin: many thanks to S. Gerischer, P. Heller, R. Wahle, S. Kempfer and P. Smeibidl for taking care of the dilution refrigerator and for making sure that the magnet was delivering its $25.9\,\mathrm{T}$.
E.F. is grateful to B. Normand for discussions and insight while shaping this work. 
K.K. would like to thank K. Kaneko (JAEA) for providing the information on sample environments at JRR-3.
\end{acknowledgments}

\medskip
\noindent
\section*{Data Availability Statement}

Raw data were generated at the BER-II research reactor at the Helmholtz-Zentrum Berlin. Derived data supporting the findings of this study are available from the corresponding author upon reasonable request.

\section*{Conflict of Interest Statement}

The authors have no conflicts to disclose.

\end{document}